\documentstyle[bezier,12pt]{article}
\newcommand{\bm}[1]{\mbox{\boldmath $#1$}}
\def\e{\mbox{e}}
\def\sn{\mbox{sn}}
\def\dn{\mbox{dn}}

\begin{document}
\title{An Example of Semiclassical Instanton-Like Scattering: (1+1)
Dimensional Sigma Model}
\author{
  V.A.Rubakov$^{a,b}$, D.T.Son$^{a}$ and P.G.Tinyakov$^{a,c}$ \\
  {\small $^{a}${\em Institute for Nuclear Research of the Russian Academy of
  Sciences,}}\\
  {\small {\em 60th October Anniversary prospect, 7a, Moscow 117312}
   (permanent address);}\\
  {\small $^{b}${\em Theoretical Physics Institute, University of Minnesota,}}
  {\small {\em Minneapolis, MN 55455;}}\\
  {\small $^{c}${\em CERN,  Geneva, CH 1211}}
  }
\maketitle

\begin{abstract}

 A solution to the classical field equations in the massless
(1+1)-dimensional $O(3)$ sigma model is found, which describes a
multi-particle instanton-like transition at high energy.
In the limit of  small number of initial
particles, the number of final particles is shown to be also small,
and the probability of the transition is suppressed by
$\exp(-2S_0)$, where $S_0$ is the instanton action. This solution,
however, does not correspond to the maximum transition probability among all
states with given number of incoming particles and  energy. Unless
the limit $g^2 n_{initial}\to0$ is exponentially sensitive to
the structure of the initial state, our results imply that well above
the sphaleron energy, the instanton-induced cross section becomes again
suppressed by the instanton exponent, and the number of final paricles is
again small.
\end{abstract}

\newpage
\section{Introduction}
  Despite considerable effort, the problem of the rate of instanton-like
transitions at high energies is still not well understood even in weakly
coupled theories. The quantitative study of this problem was initiated by
the observation\cite{Ringwald} (see also ref.\cite{Espinosa}) that in the
leading order of the pertubation theory around the instanton, the partial
cross sections rapidly grow with energy, while the total cross section is
exponential in energy \cite{MVV,Zakharov,KRT,Porrati} and naively violates
the unitarity bound at the sphaleron energy scale.

Corrections to the leading order can be
divided into three groups. The first group of corrections is entirely due to
final particles and is called soft-soft (the multiplicity in the
final state is large, and the momentum of final particles is low). These
corrections sum up into an
exponent, $\exp[F(E/E_0)/g^2]$, where $E_0$ is the
sphaleron energy, and the function $F$ contains contributions of tree
graphs only\cite{KRT,ArnoldMattis,Yaffe}. Two other groups of corrections,
namely hard-hard (that involve
only initial hard particles) and hard-soft (that involve both initial and
final
particles) contain non-trivial contributions from loops
\cite{Mueller1,Mueller2} and are harder to deal with.
There
are, however, indications\cite{Mueller2,KhlebnikovTinyakov} that they also
sum up to an exponent of the same form
as the soft-soft corrections, so one expects
that the total cross section has the following form in the regime $g\to0$,
$E/E_{0}=\mbox{fixed}$,
\begin{equation}
  \sigma=\exp\left[\frac{1}{g^2}F\left(\frac{E}{E_0}\right)\right],
  \label{sigma}
\end{equation}
The value of $F/g^2$ at
$E=0$ is of course $(-2S_{0})$ where $S_{0}$ is the instanton action;
the behavior of $F(E/E_{0})$ at $E \ll E_{0}$ is calculable by the
perturbation theory around the instanton in most models including the
standard electroweak theory.

 To evaluate the
total cross section at energies comparable to the sphaleron energy, one
has to invent a new
calculation technique.
The exponential form of eq.(\ref{sigma}) suggests
 that there may exist  semiclassical-type
methods for calculating the total cross section at all energies.
Clearly, such methods should involve a special treatment of the
non-semiclassical initial state consisting of only two hard particles,
as opposed to  the multiparticle final state whose semiclassical
treatment may be natural.

One way to gain some insight into the  problem of
instanton transitions from {\it two} particle initial state is to consider
the transitions from {\it multiparticle} states with $n=\nu/g^{2}$ initial
particles, where
$\nu$ is small enough but fixed in the limit $g^{2}\to0$ (and $E/E_0$ also
 fixed)\cite{RubakovTinyakov,Tinyakov}. In that case the problem of loop
effects does not exist, and the transition probability from a given
initial state can be evaluated, at least in principle, by entirely
semiclassical technique outlined in
 refs.\cite{RubakovTinyakov,Tinyakov,RST} (see Sect.2). This probability
indeed has the
exponential form, eq.(\ref{sigma}), where the function $F$ now depends on
the initial state and may be obtained by solving classical field equations
with certain boundary conditions on a particular contour in complex time
plane. An important issue in this procedure is the structure of
singularities of the corresponding solution in the complex time plane.

It may be quite natural that the exponent for the probability of the
transition from states with $\nu/g^2$ particles is smooth in the limit
$\nu\to0$ ({\it after} the limit $g^2\to0$ is taken) and that this limit is
independent of the precise choice of the initial states (e.g., of the
distribution of energy $E$ over $\nu/g^2$ particles). In that case the
limiting value  $F(E/E_0,\nu\to0)$ would provide an upper bound to, and may
even coincide with, the exponent for the two particle cross section.

This assumption of the independence from the limiting value $F(\nu\to0)$ on
the precise structure of initial states with $\nu/g^2$ particles, though
plausible, is by no means justified. It may be more reasonable, therefore,
to consider the "microcanonical"
probability\cite{RubakovTinyakov,Tinyakov,RST}, i.e., the
probability of the transition from a microcanonical ensemble at given
center of mass energy
$E$ and the number of particles $n=\nu/g^2$. Equivalently, one may study
{\it maximum} transition probability among all initial states of given
energy and number of particles. This probability, which is also calculable
by semiclassical methods at finite $\nu$, almost by definition gives upper
bound for the two particle cross section and is even more likely to
coincide with the latter in the limit $\nu\to0$.

Substantial support to this line of reasoning has been recently provided
by perturbative calculations about the instanton to high orders
($(E/E_0)^{16/3}$ in the electroweak theory)\cite{Mueller3}. The
conclusion of ref.\cite{Mueller3} is that in the perturbation theory about
a single instanton, the process of transition of $n_1=\nu_1/g^2$
left-moving and $n_2=\nu_2/g^2$ right-moving W-bosons into $n$ W-bosons
goes smoothly to the $2\rightarrow n$ process as $\nu_1,\nu_2\to0$.

To solve the relevant classical boundary value problem at generic values of
$E/E_0$ and $\nu=ng^2$, one almost certainly should make use of
numerical
methods. However, it is desirable to find some analytic solutions that would
provide benchmarks for numerical calculations and also indicate some
properties of the instanton-like cross sections in limiting cases. At low
energies, the approximation analogous to the dilute instanton gas is
valid, and the corresponding solution for 2d Abelian Higgs model is given
in ref.\cite{RST}. The main purpose of  this paper is to present a
solution in the 2d massless $O(3)$ sigma model that describes an
instanton-like transition in this model. One may view this model as an
extreme high energy limit, $E/E_0\gg1$, of a massive sigma
model\cite{Mottola,KRTold,Dorey}, the latter sharing most
of the  relevant properties of the
standard electroweak theory.

A peculiar feature of the massless sigma model that made it possible to
find an interesting solution is its conformal symmetry. However, precisely
this property prevents us to make contact to the low energy limit,
$E/E_0\ll1$, of the massive theory where the  transition rate is
known from the perturbation theory about the instanton\cite{KRTold,Dorey}.
Furthermore, although our solution describes {\it some} instanton-like
process, it does {\it not} correspond to any maximum (or
"microcanonical") probability in the sense of
refs.\cite{RubakovTinyakov,Tinyakov}. So, the relevance of our solution to
the two particle instanton cross section relies on the assumption of the
independence from the limit of small number of initial particles on the
precise structure of the initial state.

This paper is organized as  follows. In Sect.2 we review the
coherent state formalism in application to the study of multiparticle
transitions at fixed energy. We derive the corresponding boundary value
problem for the evaluation of the exponent for the total probability of a
transition from a given initial state to all possible final states. We
show that the sum over final states is in fact saturated by a single
coherent state whose properties are determined by the classical
solution. Inversely, for a given solution to the field equations with
appropriate singularity structure in complex time plane, one can find
the initial (and also final) state of the transition this solution
corresponds to. We also point out a specific property of  solutions
that is required for their interpretation in terms
of some "microcanonical" probability.

Before turning to the actual solution, we discuss in Sect.3 the
peculiarity of asymptotic states in (1+1) dimensional massless sigma
model. Namely, asymptotic fields moving along the light "cone" $x=t$ and
$x=-t$ are, in general, non-linear in terms of the original variables. We
describe the change of variables that linearizes the asymptotic fields (at
least in the semiclassical approximation), thus making possible their
interpretation in terms of particles.

In Sect.4 we obtain the solution
explicitly and show that it has the desired singularity structure.

Sect.5 is devoted to the interpretation of the solution in terms of an
instanton-like process. Specifically, we consider the case of small number
of initial particles, $\nu=g^{2}n\ll1$. We find that the number of final
particles is also small, $g^{2}n_{final}\ll 1$
 (but non-analytic in $\nu$), and $g^{2}n_{final}$
 vanishes in the
limit $\nu\to0$. The transition probability in this limit tends to the
instanton value. Under the assumption of the independence from the limit of
small number of initial particles on the structure of initial state, this
implies that the scattering at very high energies again becomes suppressed
precisely by the instanton exponent. We show, however, that our solution
cannot be interpreted as the configuration corresponding to the maximal
transition probability among initial states with given number of initial
particles; this makes the indication  on the suppression of the
instanton-like  cross section at extremely high energies by precisely the
instanton exponent less clear.

Sect.6 contains concluding remarks.

\section{Classical solutions and multiparticle scattering at fixed energy}
  In the spirit of refs.\cite{RubakovTinyakov,Tinyakov}, let us consider a sum
\begin{equation}
  \sigma(\{ a_{\bf k}\})=\sum_b \mid \langle \{b_{\bf k}\}\mid
     SP_E\mid \{a_{\bf k}\} \rangle \mid^2,
  \label{sum}
\end{equation}
where $P_E$ is a projector onto a subspace of definite energy $E$,
$S$ is the instanton $S$-matrix and
$~~\mid \{a_{\bf k}\} \rangle $, $~~\mid \{b_{\bf k}\} \rangle ~~$ are coherent
states.
This sum is the probability of a transition  at fixed energy $E$ from
some coherent state $\mid \{a_{\bf k}\} \rangle $
(projected onto this energy) to all possible final states. We consider the
system in its center of mass frame and will not discuss in what follows the
constraints due to the total spatial momentum.

 Making use of
the functional integral representation for the $S$-matrix elements, one writes
the transition probability
 in the following integral form\cite{KRTperiodic,RubakovTinyakov}
\[ \sigma= \int
  db^*_{\bf k}db_{\bf k}d\xi d\xi'd\phi(x)d\phi'(x) \exp\big(-\int d\bm{k}
  b^*_{\bf k}b_{\bf k}-iE(\xi-\xi')+
\]
\begin{equation}
  B_i(a_{\bf k}\e^{i\omega_k\xi}, \phi_i)+B_f(b^*_{\bf k}, \phi_f) +iS(\phi)+
  B_i^*(a^*_{\bf k}\e^{-i\omega_k\xi'}, \phi_i')+B_f^*(b_{\bf k},
  \phi_f') -iS(\phi')\big)
  \label{int},
\end{equation}
where $\phi$ is the condensed notation for the (real) bosonic fields of the
theory,
$S(\phi)$ is the action, $B_i$ and $B_f$ stand for boundary terms and are
merely
logarithms of the wave functions of coherent states in the $\phi$
representation,
  \[
  B_i(a_{\bf k},\phi_i) =
  - \frac{1}{2} \int\! d\bm{k} \, a_{\bf k} a_{\bf -k} e^{-2i\omega_kT_i}
  - \frac{1}{2} \int\! d\bm{k} \omega_{\bf k}
  \phi_i(\bm{k})\phi_i(-\bm{k})
  \]
  \[
  + \int\! d\bm{k} \sqrt{2\omega_{\bf k}}\, e^{-i\omega_kT_i}
  a_{\bf k} \phi_i(\bm{k}) \,
  \]
  \[
  B_f(b^*_{\bf k},\phi_f) =
  - \frac{1}{2} \int\! d\bm{k} \, b^*_{\bf k} b^*_{\bf -k}e^{2i\omega_kT_f}
  - \frac{1}{2}\int\! d\bm{k} \omega_{\bf k}
  \phi_f(\bm{k}) \phi_f(-\bm{k})
  \]
  \begin{equation}
  + \int\! d\bm{k} \sqrt{2\omega_{\bf k}}e^{i\omega_kT_f}
  b^*_{\bf k} \phi_f(-\bm{k}) \; .
  \label{B}
  \end{equation}
In the above formulae, $T_i$ and $T_f$ are initial and final times,
$\phi_i,\phi_f=\phi (T_{i}),\phi (T_{f})$ and the
limit $T_i\to-\infty$, $T_f\to+\infty$ is assumed.

At $E\sim 1/g^{2}$ and $a_{\bf k} \sim 1/g$, that corresponds to the
regime
\[
 g^2\to0~, ~~~~\nu = ng^{2}=\mbox{fixed}
\]
 discussed throughout this paper,
the integral in eq.(\ref{int}) can be evaluated in the saddle-point
approximation. Varying the exponent in eq.(\ref{int}) with respect to
$\phi$, $\phi'$ at $T_{i}<t<T_{f}$, we obtain  that the saddle
point
configuration for $\phi$, $\phi'$ should be a solution to the field equations
$\delta S/\delta\phi=\delta S/\delta\phi'=0$, while variation with respect to
$b^*_{\bf k}$, $b_{\bf k}$ and boundary values of $\phi$ and $\phi'$ leads
to six equations which relate the boundary values of $\phi(x)$,
$\phi'(x)$ and their time derivatives to $a_{\bf k}$,
$a^*_{\bf k}$, $b_{\bf k}$, $b^*_{\bf k}$. In particular, the integration
over $b$, $b^*$, $\phi_{f}$ and $\phi_{f}'$ results in the boundary conditions
\begin{equation}
\phi'(x)=\phi(x),~~~~~~~~~~
\dot{\phi}'(x)=\dot{\phi}(x)
\label{same}
\end{equation}
at $t\to +\infty$, and
\[
  i\dot{\phi}_f(\bm{k})-\omega_{\bf k}\phi_f(\bm{k})+
  \sqrt{2\omega_{\bf k}}b^*_{-{\bf k}}\e^{i\omega_kT_f}=0
\]
\begin{equation}
  -i\dot{\phi}_f(\bm{k})-\omega_{\bf k}\phi_f(\bm{k})+
  \sqrt{2\omega_{\bf k}}b_{\bf k}\e^{-i\omega_kT_f}=0
\label{b's}
\end{equation}
Eq.(\ref{same}) implies that
\[
  \phi'(x)=\phi(x)
\]
everywhere in space-time, i.e., we are dealing with a single solution to
the field equations. Eqs.(\ref{b's}) determine the saddle point values of
the coherent state variables corresponding to the final state.

  The saddle point equations at the initial time are
\[
  i\dot{\phi}_i(\bm {k})-\omega_{\bf k}\phi_i(\bm {k})+
  \sqrt{2\omega_{\bf k}}a^{*}_{-\bf k}e^{i\omega_kT_i-i\omega_k\xi'}
  =0
\]
\begin{equation}
  -i\dot{\phi}_i(\bm{k})-\omega_{\bf k}\phi_i(\bm{k})+
  \sqrt{2\omega_{\bf k}}a_{\bf k}\e^{-i\omega_kT_i+i\omega_k\xi}=0,
\label{6}
\end{equation}
Notice that we keep the initial state, i.e., the amplitudes $a_{\bf k}$
and $a^{*}_{\bf k}$, fixed, so that eqs.(\ref{6}) are to be considered as
the boundary conditions for the field $\phi$ at large negative time.

Finally, the variation with respect to $\xi$ and $\xi'$ ( in fact, the
integral over $(\xi+\xi')$ is trivial due to the time translation
 symmetry ) relates  the
energy $E$ to other parameters of the solution. This relation is  very
simple  and will be explicit later.

The above boundary conditions can be converted
into more convenient form. First, we assume that at the initial and final
times $T_i$ and $T_f$,
the field
$\phi$ becomes free and can be written as a
superposition of plane waves.
In particular, at large positive time
\[
  \phi(\bm{k},t)=\frac{1}{\sqrt{2\omega_{\bf k}}}
  \big(g_{\bf k}\e^{-i\omega_kt}+\bar{g}_{-{\bf k}}
  \e^{i\omega_kt}\big),~~~~~t\to+\infty.
\]
  Substituting this asymptotics into eq.(\ref{b's}) we obtain
\begin{equation}
  b_{\bf k}=g_{\bf k},~~~~~b^*_{\bf k}=\bar{g}_{\bf k}.
\label{BK}
\end{equation}
Notice that at this point we cannot insist that the saddle point field is
real, so in general $\bar{g}_{\bf k}$ need not be complex conjugate to
$g_{\bf k}$.

Second, the dependence of the integrand in eq.(\ref{int}) on $(\xi+\xi')$
is trivial, so we can set $\xi'=-\xi$. Furthermore, the real part of the
saddle point value of $\xi$ can be removed by the time translation, and we
can choose
\[
  \xi=iT
\]
Consider now the contour in complex time plane shown in fig.1. On this
contour, the field at early time,
\[
 t=iT+t',~~~~~ t'\to -\infty
\]
has the following form,
\begin{equation}
  \phi(\bm{k},t)=\frac{1}{\sqrt{2\omega_{\bf k}}}
  \big(f_{\bf k}\e^{-i\omega_kt'}+
  \bar{f}_{-{\bf k}}\e^{i\omega_kt'}\big)
\label{freq}
\end{equation}
where  $\bar{f}_{\bf k}$ need not be complex conjugate to $f_{\bf k}$. The
second of eqs.(\ref{6}) gives then
\begin{equation}
  f_{\bf k}=a_{\bf k}
\label{AK}
\end{equation}
Analogously, the first boundary condition in eq.(\ref{6}) is conveniently
formulated at $\mbox{Im}~t=-T$, where the field is
\[
  \phi(\bm{k},t'')=\frac{1}{\sqrt{2\omega_{\bf k}}}
  \big(h_{\bf k}\e^{-i\omega_kt''}+
  \bar{h}_{-{\bf k}}\e^{i\omega_kt''}\big)
\]
\[
  t=-iT+t'',~~~~~~t''\to -\infty
\]
The boundary condition reads
\[
  \bar{h}_{\bf k}=a^*_{\bf k}
\]
Since $a^*_{\bf k}$ ans $a_{\bf k}$ are complex conjugate to each other,
the latter condition is equivalent to the requirement that the solution
$\phi(x)$ is real on the real (minkowskian) part of the contour, $C$ in
fig.1, provided the solution is unique. This means, in particular,
that the saddle point values of
$b^*_{\bf k}$ and $b_{\bf k}$ are complex conjugate to each other, i.e.,
the sum over final states in eq.(\ref{sum}) is saturated by a single
coherent state.

To summarize, the transition probability $\sigma(\{a_{\bf k}\})$ is
determined, in the semiclassical approximation, by a solution to the
classical field equations on the contour of fig.1, obeying the following
boundary conditions: (i) the field must be real in minkowskian part of the
contour; (ii) the negative frequency ("anti-Feynman") part of the field
at large negative time (part $A$ in fig.1) must be determined by the
amplitudes $a_{\bf k}$ via eqs.(\ref{freq}),(\ref{AK}). The final state is
then given by the asymptotics of the solution at large positive time by
eq.(\ref{BK}).

 Finally, one can see that the saddle point value of $T$ is
determined by
the requirement that the classical energy of the solution is equal to $E$.

A few remarks are in order. First, the solution need not be real on the
euclidean part and/or on the part $(AB)$ of the contour of fig.1.
Furthermore, the solutions real everywhere on this contour do not
correspond to the case when the number of initial particles is small
\cite{KRTperiodic,RubakovTinyakov}. So, most interesting solutions even in
theories with hermitean fields are complex. Second, interesting
solutions, though smooth at our contour, should have singularities in the
complex time plane between the real axis and the line $(AB)$, otherwise they
would correspond to classically allowed scattering.

Finally, instead of considering transitions from a given initial state, on
may study the "microcanonical" probability\cite{RubakovTinyakov,Tinyakov},
\begin{equation}
  \sigma_{E,D}=\sum_{i,f}\mid \langle f\mid SP_E P_D\mid i \rangle \mid^2
\label{SED}
\end{equation}
where the sum runs over all final {\it and} initial states and $P_D$ is
the projector onto an eigenspace of some operator
\[
  D=\int \! d\bm{k}\Delta_{\bf k}A^{\dagger}_{\bf k}A_{\bf k}
\]
with real $\Delta_{\bf k}$ (an obvious choice would be $\Delta_{\bf k}=1$
that would correspond to fixing the number of initial particles). Here
$A^{\dagger}_{\bf k}$ and $A_{\bf k}$ are creation and annihilation operators
for initial
particles. It follows from refs.\cite{RubakovTinyakov,Tinyakov,RST} that the
sum in
eq.(\ref{SED}) is also saturated by a single initial state which therefore
has the {\it maximum} transition probability among the states with given
energy $E$ and the value of $D$. Instead of eq.(\ref{AK}), the boundary
condition for the corresponding saddle point field would be
\begin{equation}
  f_{\bf k}^{*}= e^{-\theta\Delta_{k}}\bar{f}_{\bf k}
\label{maximum}
\end{equation}
in the region $(A)$ of the contour of fig.1, where $\theta$ is some real
constant and $f_{\bf k}^{*}$ is complex conjugate to $f_{\bf k}$.
 Notice that in this case the ratio of positive ("Feynman") and
negative ("anti-Feynman") frequency parts of the saddle point field at
$t=iT+t'$, $t'\to -\infty$, should be real and positive.

Let us now come back to the discussion of the transitions from a fixed
initial state. The above arguments show that every solution to the field
equations, that is real on real time axis and has an appropriate singularity
structure in complex time plane, corresponds to such a transition from
{\it some} initial state. The amplitudes characterising the initial state,
$\{a_{\bf k}\}$, are determined by the negative frequency part of the
solution in the region $A$ of the contour via eq.(\ref{AK}), while the
most probable final state is determined by the behavior of the solution
at large real time by eq.(\ref{BK}). There is, however, an apparent
ambiguity in the determination of the initial state of the process described
by a given
solution. Namely, one can shift the line $AB$ of this contour upwards or
downwards (until this line reaches the singularity) thus changing the
asymptotics of the field along the contour at negative $\mbox{Re}~t$. This
would
induce a change in the amplitudes $a_{\bf k}$ of the initial state. In
other words, the choice of the value of $T$ for a given solution has
apparent arbitrariness, that translates into the apparent arbitrariness of
the initial state.

Let us see that this ambiguity is in fact irrelevant. The key observation
is that
the projections of the coherent states of the form $\mid a_{\bf
k}\e^{\omega_k\eta} \rangle $ with various $\eta$ onto the subspace of the
hamiltonian $H$  with energy $E$ differ only by normalization:

\[ P_E\mid a_{\bf k}\e^{\omega_k\eta} \rangle = \int dt\e^{i(H-E)t}\mid a_{\bf
  k}\e^{\omega_k\eta} \rangle =
\]
\[ =\int dt\e^{-iEt}\mid
  a_{\bf k}\e^{i\omega_kt+\omega_k\eta} \rangle = \int
dt'\e^{-iE(t'+i\eta)}\mid
  a_{\bf k}\e^{i\omega_kt'} \rangle = e^{E\eta}P_E\mid a_{\bf k} \rangle
\]
Suppose now that there are no
singularities of the field $\phi$ in the region $T<\mbox{Im}~t<T'$.
In this case the negative frequency components $f_{\bf k}$ and $f'_{\bf k}$
of  the  solution on the lines $\mbox{Im}~t=T$ and
$\mbox{Im}~t=T'$, respectively,
are related to each other as follows,
\[
  f'_{\bf k}=f_{\bf k}\e^{\omega_k\eta},~\eta=T'-T
\]
The same relation holds for the amplitudes characterising initial states,
so their projections onto the subspace of fixed energy $E$ are essentially the
same.

To find the most convenient choice of the contour (i.e., the most
convenient value of the parameter $T$), let us first fix $T$ in an
arbitrary way and
calculate the average of some bilinear operator
\[
     O=\int d\bm{k}F(\bm{k})A_{\bf k}^{\dag} A_{\bf k}
\]
over the state  $P_E\mid a_{\bf k} \rangle $ determined by the solution on the
contour that starts at $\mbox{Im}~t=T$. We have
\[
      \langle a_{\bf k}\mid P_EOP_E\mid a_{\bf k} \rangle =
\]
\[
     \int dtdt'\e^{-iE(t-t')}
      \langle a_{\bf k}\e^{i\omega_kt'}\mid\int d\bm{k}F(\bm{k})
     A_{\bf k}^{\dag} A_{\bf k}\mid a_{\bf k}\e^{i\omega_kt} \rangle =
\]
\begin{equation}
     \int dtdt'\exp\left(-iE(t-t')+\int d\bm{k} f^*_{\bf k}f_{\bf k}
     \e^{i\omega_k(t-t')}\right)
     \int d\bm{k}F(\bm{k})f^*_{\bf k}f_{\bf k}
     \e^{i\omega_k(t-t')}
     \label{norm}.
\end{equation}
where $f^*_{\bf k}$ is {\it complex conjugate} to $f_{\bf k}$ (notice that
$f^*_{\bf k}$ are {\it not} equal to the positive  frequency components of
the solution, $\bar{f}_{\bf k}$).
The integral can be evaluated in the saddle-point approximation.The result is
\[
      \langle O \rangle =\int d\bm{k}F(\bm{k})f^*_{\bf k}f_{\bf k}\e^{-\omega
T_0},
\]
where $T_0$ is detrmined by the following relation,
\begin{equation}
     E=\int d\bm{k}\omega_{\bf k}f^*_{\bf k}f_{\bf k}\e^{-\omega_k T_0},
     \label{E}
\end{equation}
(the factor $\exp(ET_0+\int d\bm{k}f^*_{\bf k}f_{\bf k}\e^{-\omega T_0})$
that appears in eq.(\ref{norm}) at the saddle-point  is merely
the norm of the state $P_E\mid a_{\bf k} \rangle $). If instead of
$a_{\bf k}=f_{\bf k}$ we used $\alpha_{\bf k}=f_{\bf k}\e^{-\omega_k T_0/2}$
(so that $P_E\mid a_{\bf k} \rangle $ and $P_E\mid \alpha_{\bf k} \rangle $ are
the same
states with different
normalizations), then the average of $O$ would have a simple form,
\[
      \langle O \rangle =\int d\bm{k}F(\bm{k})\alpha^*_{\bf k}\alpha_{\bf k}.
\]
Note that the average energy of the coherent state $\mid \alpha_{\bf k} \rangle
$ is
equal to $E$ (eq.(\ref{E})). If there are no singularities in the region
$T-T_0/2<\mbox{Im}~t<T$, the positive frequency components of the
solution calculated
at $t=i(T-T_0/2)+t'$, $t'\to-\infty$ are exactly
equal to $\alpha_{\bf k}$.
So, in some cases it is
possible to choose $T$ in the most convenient way, namely, to satisfy the
condition
 \[
     E=\int d{\bf k}\omega_{\bf k}f^*_{\bf k}f_{\bf k}.
\]
With this choice of the contour, the averages of operators over the
initial state (say, the number of particles) are expressed through
$(a_{\bf k},a^*_{\bf k})=(f_{\bf k},f^*_{\bf k})$ in a standard way.

So, every solution that is real on the real time axis and has correct
singularity structure in the complex time plane (so that one can
really draw a contour of fig.1 at which the solution is non-singular)
can be
considered as a
configuration that dominates some process at fixed energy. The
semiclassical probability
of this process can be found by substituing the saddle point values
of the integration variables
into eq.(\ref{int}) and dividing the result by the norm of the state
$P_E\mid a_{\bf k} \rangle $, which is equal to $\exp(\int d\bm{k}
a^*_{\bf k}a_{\bf k})$  for the
contour with ''correct'' $T$. One obtains for the semiclassical
 probability
\begin{equation}
  \sigma=\exp\left(2ET-2\mbox{Im}~S(\phi)+\int d{\bf k}(\bar{f}_{\bf k}-
  f^*_{\bf k})f_{\bf k}\right)
  \label{prob}.
\end{equation}

If, furthermore, at large negative time (part $A$ of the contour)
 positive and negative frequency components for a given solution
are complex conjugate to each other up to a real
and positive
factor for all $\bm{k}$, then this solution can be interpreted as
describing {\it some} microcanonical probability: indeed, in that case one
can choose a suitable form of $\Delta_{\bf k}$ to satisfy
eq.(\ref{maximum}). We shall see that this property is absent for the
sigma model solution of Sect.4.

 To apply
this
formalism to the (1+1) dimensional sigma model we have, however, to be able
 to describe its asymptotic states.

\section{Asymptotic states in $O(3)$ sigma model}
  Consider the massless non-linear $O(3)$ sigma model with the action
\[
  S=\frac{1}{2g^2}\int d^2x(\partial_\mu{\bf n})^2
  =\frac{2}{g^2}\int d^2x\frac{\partial_\mu\bar{w}\partial_\mu w}
  {(1+\bar{w}w)^2},
\]
\[
 {\bf n}^2=1
\]
where $\bar{w}$ and $w$ are fields in the projective representation,
\begin{equation}
  n_1=\frac{w+\bar{w}}{1+\bar{w}w},~~~~~
  n_2=\frac{1}{i}\frac{w-\bar{w}}{1+\bar{w}w},~~~~~
  n_3=\frac{1-\bar{w}w}{1+\bar{w}w}.
\label{nw}
\end{equation}
A peculiar feature of this model is that the fields cannot be considered
free in
asymptotic regions $t\to\pm\infty$. Indeed, any field configuration of the
form $w=w(x_+)$ or $w=w(x_-)$ where $x_+=x+t$, $x_-=x-t$,
satisfies the field equations. Its  amplitude
need not be small, and the non-linearity persists even at large
positive and/or negative
$t$. The  interpretation of classical solutions in terms of particles
requires a change of variables.

  Let us consider only such solutions that in the limit $t\to\infty$ split
into two non-linear wave packets moving along two parts of the light "cone"
$x=t$ and $x=-t$,
\[
 w(x,t)=w_+(x_+)+w_-(x_-).
     \label{as}
\]
At large time the two wave packets do not interact, so it is natural
that left moving and right moving waves  can be considered separately. Let,
for example,
 $w=w(x_-)$ be right moving quantum field. Let us
examine the generator of the translation
along the $x_-$ direction,
\[
	P_-=\frac{4}{g^2}\int d^2x\frac{\partial_-\bar{w}\partial_-w}
	{(1+\bar{w}w)^2},
\]
It has the following property,
\[
  [P_-,w(x_-)]=i\partial_-w(x_-),
\]
\[
  [P_-,\bar{w}(x_-)]=i\partial_-\bar{w}(x_-),
\]
where we have used the notation $\partial_-=\partial/\partial x_-$.
Let us now define new fields $\Phi(x_-)$ and $\bar{\Phi}(x_-)$ by
the following relations, which are unambiguous at least in the
semiclassical limit,
 \begin{equation}
  \partial_-\Phi=\frac{\sqrt{2}}{g}\frac{\partial_-w}{1+\bar{w}w},~~~
  \partial_-\bar{\Phi}=\frac{\sqrt{2}}{g}\frac{\partial_-\bar{w}}{1+\bar{w}w},
\label{FI}
\end{equation}
Then the generator $P_-$ becomes
\[
	P_-=2\int d^2x\partial_-\bar{\Phi}\partial_-\Phi,
\]
while its commutators with $\Phi$ and $\bar{\Phi}$ have the same
functional form as with $w$ and $\bar{w}$,
\[
  [P_-,\Phi(x_-)]=i\partial_-\Phi(x_-),
\]
\[
  [P_-,\bar{\Phi}(x_-)]=i\partial_-\bar{\Phi}(x_-).
\]
To ensure these commutation ralations, it is sufficient to require
the following commutation ralations for $\Phi$ and $\bar{\Phi}$,
\[
     [\Phi(x_-),\Phi(y_-)]=[\bar{\Phi}(x_-),\bar{\Phi}(y_-)]=0,
\]
\begin{equation}
     [\Phi(x_-),\bar{\Phi}(y_-)]=\frac{i}{4}\epsilon(x_- - y_-).
     \label{cr}
\end{equation}
In the Fourier representation,
\[
     \Phi(x_-)=\int_0^\infty\frac{dk}{\sqrt{2\pi}\sqrt{2k}}
     \big(\hat{\alpha}_k\e^{ikx_-}+\hat{\beta}_k^{\dag}\e^{-ikx_-}\big),
\]
\[
     \bar{\Phi}(x_-)=\int_0^\infty\frac{dk}{\sqrt{2\pi}\sqrt{2k}}
     \big(\hat{\beta}_k\e^{ikx_-}+\hat{\alpha}_k^{\dag}\e^{-ikx_-}\big),
\]
the commutation relations (\ref{cr}) imply that the operators
$\hat{\alpha}_k$,
$\hat{\alpha}_k^{\dag}$, $\hat{\beta}_k$, $\hat{\beta}_k^{\dag}$ can be
interpreted as
annihilation and creation operators for particles and anti-particles,
respectively,

\[
     [\hat{\alpha}_k, \hat{\alpha}_{k'}^{\dag}]=\delta(k-k'),~~~
     [\hat{\beta}_k, \hat{\beta}_{k'}^{\dag}]=\delta(k-k'),
\]
and other commutators vanish.
Thus, the change of variables, eq.(\ref{FI}), enables us to interprete the
outgoing right moving fields in terms of  particles and antiparticles
of spatial momentum $k>0$.
 Quantizing the left-moving waves we
describe particles and antiparticles with negative momenta in the same way.
Note that our interpretation
is essentially based on the separation of two wave packets in the
asymptotic region (eq.(\ref{as})).

     Once we have defined annihilation and creation operators, we can
construct coherent states in the standard way,
\[
     \hat{\alpha}_k\mid \alpha_k \rangle =\alpha_k\mid \alpha_k \rangle ,~~~
     \hat{\beta}_k\mid \beta_k \rangle =\beta_k\mid \beta_k \rangle ,
\]
and make use of the coherent-state formalism for calculating
transition probabilities at
fixed energy in the sigma model.

\section{The classical solution}

Our approach to obtain the classical solutions that are real (in the sense
that $\bar{w}$ is complex conjugate to $w$) in Minkowski time and have
singularities in the complex time plane is to start from {\it euclidean}
space-time
 $(x, \tau)$ and search for real solutions that have a turning point,
 $\partial_\tau w(x)=0$ for all $x$ at $\tau=0$.
The reality of the solutions in the euclidean domain and
the existence of the turning point then guarantee that their analytical
continuation to the complex time plane is real at Minkowski time. Since we
are interested only in a finite region of euclidean time, $0<t<T$,
the solutions may (and, in fact, should) have singularities outside this
region.

Since the action of the sigma model is conformally invariant, the field
equations do not change under the following conformal mapping,
\begin{equation}
  z'=x'+i\tau'=\frac{z-i}{z+i}=\frac{x+i\tau-i}{x+i\tau+i}~,
  \label{trans}
\end{equation}
that tranforms the upper half-plane into a disc of unit radius, centered
at $z'=0$, in which we solve the field equations. [There is no intrinsic
mass scale in the model at the classical level, so we do not introduce a
length scale into the solution. We will be interested in dimensionless
quantities only.]
The existence of the turning point in the original variables is equivalent,
in terms of the
 new
variables, to the boundary condition that the normal derivative of $w$ at the
circle of unit radius that bounds the disc, is equal to zero. If we introduce
the polar coordinates $\rho$ and $\varphi$ on the $z'$-plane,
\[
  \rho=\mid z'\mid=\left(\frac{x^2+(\tau-1)^2}{x^2+(\tau+1)^2}\right)^{1/2}
  ,~~~
  \e^{i\varphi}=\left(\frac{(x-i)^2+\tau^2}{(x+i)^2+\tau^2}\right)^{1/2},
\]
then the boundary condition becomes $\partial_\rho w\mid_{\rho=1}=0$.

We  search for $O(2)$ symmetric solutions,
\[
w=f(\rho)\e^{i\varphi},~~~~~~~~~~~~~ \bar{w}=f(\rho)\e^{-i\varphi}
\]
The action for these configurations has the form
\[
  S=\frac{4\pi}{g^2}\int_0^1\rho d\rho\frac{1}{(1+f^2)^2}
  \left[\left(\frac{\partial f}{\partial \rho}\right)^2+\frac{f^2}{\rho^2}
  \right].
\]
Making the change of variables\cite{British}
$f=\tan(\psi/4)$, $\rho=\exp(\zeta)$ one obtains
\[
  S=\frac{\pi}{2g^2}\int_0^\infty d\zeta\left[\frac{1}{2}
  \left(\frac{\partial\psi}{\partial\zeta}\right)^2
  +(1-\cos\psi)\right].
\]
The problem  is now reduced to the
 mechanical problem of a particle moving in the sin-Gordon potential
$U(\zeta)=-(1-\cos\psi)$ with the initial condition
$\partial\psi/\partial\zeta=0$ at $\zeta=0$ (Fig.2). There exists a set of
solutions depending on a single parameter $\psi_0$ which is the initial
coordinate of the particle. It is obvious that the solution $\psi(\zeta)$
is  periodic. The oscillatory behavior of $\psi(\zeta)$ means
that the point $\rho=0$ ($\zeta=-\infty$) is an essential singularity. In
the initial $z$-plane this essential singularity appears at
$x=0$, $\tau=1$. The existence of essential singularities in the  euclidean
time axis
may, in fact,  have been expected for
configurations describing processes in which the number of initial and
final particles are not equal to each other\cite{RST}.

The explicit form of the solution is
 \begin{equation}
  \psi=2\arccos[-k_0\sn(\zeta-K, k_0)],
  \label{arccos}
\end{equation}
where $k_0=\cos(\psi_0/2)$, $\sn$ is the Jacobi elliptic sine, and
\[
  K=\int_0^{\pi/2}\frac{dx}{\sqrt{1-k_0^2\sin^2x}},
\]
is a quarter of the period of function $\psi(\zeta)$.
The fields $w$ and $\bar{w}$ are then,
\begin{equation}
  w=\left(\frac{1+k_0\sn(\zeta-K)}{1-k_0\sn(\zeta-K)}\right)^{1/2}
  \e^{i\varphi},~~~~~
  \bar{w}=\left(\frac{1+k_0\sn(\zeta-K)}{1-k_0\sn(\zeta-K)}\right)^{1/2}
  \e^{-i\varphi},
  \label{solution}
\end{equation}
where $\zeta$ and $\varphi$ must be understood as functions of $x$
and $\tau$,
\begin{equation}
  \zeta=\ln\rho=\frac{1}{2}\ln\frac{x^2+(\tau+1)^2}{x^2+(\tau-1)^2}
,~~~~~
  \varphi=\frac{i}{2}\ln\frac{(x+i)^2+\tau^2}{(x-i)^2+\tau^2}
	\label{zp}.
\end{equation}

Now we can continue analytically the solution to complex time plane. This
is done by setting $\tau=-it$ where $t$ will finally run along
the contour of fig.1. Eq.(\ref{zp}) transforms into
\begin{equation}
  \zeta(x,t)=\frac{1}{2}\ln\frac{x^2+(1-it)^2}{x^2+(1+it)^2}
,~~~~~
  \varphi=\frac{i}{2}\ln\frac{(x+i)^2-t^2}{(x-i)^2-t^2}
	\label{zpt}.
\end{equation}
When $t$ is complex, $\varphi$ and $\zeta$ are, in general, also
complex, so the fields $w$ and $\bar{w}$ are not complex conjugate to
each other.

Let us now turn to the discussion of the singularities of our solution in
complex time. There exist essential singularities at
\[
 t=\pm x +i
\]
(we concentrate on the region $\mbox{Im}~t>0$). Their positions make a
"cross" in the three dimensional space $(\mbox{Re}~t,\mbox{Im}~t,x)$, as
shown in fig.3. Clearly, we should choose $T<1$ for the contour of fig.1
not to intersect this cross.

One can see that apart from these essential singularities, the fields $w$
and $\bar{w}$ have poles at complex $t$. However, at these points the
original fields $\bf{n}$ and other dynamical variables (like energy
density) are regular. Therefore, these poles have no physical
significance.

On the other hand, the fields $\bf{n}$ and other dynamical quantities are
singular at the points where
\[
 1+\bar{w}w=0
\]
i.e., where the transformation
 (\ref{nw}) is singular. At these points the dynamical variables have
poles, and the positions of these poles, $t=t(x)$, make (non-planar)
curves in the space $(\mbox{Re}~t,\mbox{Im}~t,x)$, as shown schematically
in fig.3.

We will be interested mostly in the case
\[
\psi_0\ll 1
\]
which, as will be shown in Sect.5, corresponds to small number of initial
particles. Let us describe the positions of the physical singularities in
that case.
 Making use of the
explicit form of the solution, we write
\[
  1+\bar{w}w=\frac{2}{1-k_0\sn(\zeta-K)}~,
\]
which is zero when $\sn(\zeta-K)$ is infinite, or $\zeta=(2n+1)K+(2m+1)iK'$,
where
 \[
  K'=\int_0^{\pi/2}\frac{dx}{\sqrt{1-k_1^2\sin^2x}},
\]
\[
	k_1^2=1-k_0^2=\sin^2\frac{\psi_0}{2}.
\]
At small $\psi_0$, the equation $\zeta(x,t)=(2n+1)K+(2m+1)iK'$, where
$\zeta(x,t)$
is given by eq.(\ref{zpt}),
 has solutions only at $m=-1$ and
$n\geq0$.
The singularity curves
\[
	t=t_n(x),~~~~~n=0,1,2,\ldots
\]
 have $\mbox{Im}~t$ close to
1, and the larger is $n$, the closer is the curve $t=t_n(x)$ to the cross
$t=\pm x+i$ (two of these curves are shown in fig.3; our solution is
symmetric under the transformation $t\to-t$, so there are also singularity
curves at
$\mbox{Re}~t>0$, not shown in fig.3). Asymptotically, these curves
run parallel to the light "cone", and the behavior of $t_n(x)$ at large
$\mid x\mid$ is
 \[
	t_n=-\mid
	x\mid+i\left(1-2\left(\frac{k_1}{4}\right)^{2(2n+1)}+\cdots\right)
  ~~~\mid x\mid\to\infty.
\]

	The integration contour of fig.1 can be imagined as a ladder step of
height $T$ in the
three-dimensional picture of fig.3. By varying $T$ we can place
the upper part of the step, $(\mbox{Re}~t<0,~\mbox{Im}~t=T)$, above
different number of the singularity curves. To avoid crossing the
singularities, one should in general consider curved surfaces, not just
ladder steps, but this is unnecessary in our case for the surfaces with
exactly
{\it one}
singularity line under them. In what follows we will consider only such
surfaces, i.e., we will choose
\[
    1-k_1^2/8<T(\infty)<1-k_1^6/2^{11} .
\]
This choice corresponds to one-instanton transitions, while surfaces with
more than one singularity curves under them would most likely correspond
to multi-instanton processes.

The  solution is explicitly real (in the sense that $\bar{w}$ is complex
conjugate to $w$) in Minkowski
time. This is clear from the following equivalent form of the solution,
\begin{equation}
  w=\left(\frac{\dn (i\zeta, k_1)-k_0}{\dn (i\zeta, k_1)+k_0}
  \right)^{1/2}\e^{i\varphi},
\label{Form}
\end{equation}
and the observation that at real $t$, the variables $\zeta$ and $\varphi$
determined by eq.(\ref{zpt}) are pure imaginary and real, respectively.

To calculate the imaginary part of the action and the topological charge $Q$,
we notice that
for every  $x$, the
integral
\begin{equation}
  S(x)=\frac{2}{g^2}\int_C dt\frac{\partial_{\mu}\bar{w(x,t)}
  \partial_{\mu}w(x,t)}{(1+\bar{w}(x,t)w(x,t))^2},
  \label{Sx}
\end{equation}
on the integration contour of fig.1 is a sum of the  same integral over
the real time axis, and the contribution from the pole $1+\bar{w}w=0$.
On the real time axis the integral is real, so the imaginary part comes
from the pole.
[The integral for the topological charge over the real time axis is zero
because of the invariance of the solution under $t\to -t$.] Then the
calculation is straightforward, and the result is
\[
 \mbox{Im}~S=\int dxS(x)=\frac{4\pi}{g^2}
\]
In the same way we find
\[
   Q=1
\]
Thus, the action is exactly equal to the instanton action for all
$\psi_0$, and our solution indeed corresponds to a one-instanton transition.

At large $\mid x\mid$ and fixed $t$, the fields tend to
\[
   w=\bar{w}=w_0
\]
where
\begin{equation}
 w_0=\tan (\psi_0/4)
\label{w0}
\end{equation}
Therefore, our configuration is a fluctuation about a vacuum with
non-vanishing expectation values of $w$ and $\bar{w}$, i.e., about a
vacuum with ${\bf n} \neq (0,0,1)$. It is convenient to perform an $O(3)$
rotation
in the ${\bf n}$-space
so that the vacuum expectation value  becomes ${\bf n}=(0,0,1)$.
In terms of the
 projective
variables this rotation reads
\begin{equation}
  w'=\frac{w-w_0}{1+w_0w}~,~~~~\bar{w}'=\frac{\bar{w}-w_0}{1+w_0\bar{w}}~.
  \label{rot}
\end{equation}
The fields $w'$ amd $\bar{w}'$ are then the fluctuations about their zero
vacuum expectation values.

Let us discuss the solution in the case $\psi_0\ll 1$ in more detail.
In that case the singularity line is near the plane $\mbox{Im}~t=1$. Well
below this plane,
we make use of  the following
approximation for $\dn(i\zeta, k_1)$ at small $k_1$ \cite{GrRyzh},
\begin{equation}
  \dn(i\zeta, k_1)=1+\frac{k_1^2}{2}\sinh^2\zeta,
  \label{appr}
\end{equation}
so that to the leading order in $\psi_0$ we have
\[
  w'=-iw_0\left(\frac{1}{x+t+i} + \frac{1}{x-t+i}\right),
\]
\begin{equation}
  \bar{w}'=iw_0\left(\frac{1}{x+t-i}+\frac{1}{x-t-i}\right),
\label{asymp}
\end{equation}
\[
   w_0=\frac{1}{4}\psi_0
\]
In the euclidean domain, $t=i\tau$, this field configuration is that of
widely separated  instanton and anti-instanton of sizes $w_0\ll 1$,
 sitting at
$x=0$ and $\tau=+1$ and $\tau=-1$, respectively. This property may have
been expected on the basis of the results of ref.\cite{RST}. At
$(1-\tau)\sim w_0^2$ we cannot, however, use the approximation (\ref{appr})
and the solution differs from the instanton~-~anti-instanton pair
considerably.

To end up this section, let us point out that our solution is very similar
to the L\"uscher-Schechter solution in $(3+1)$ pure Yang-Mills
theory\cite{Luscher,Schechter}. In fact, we could generate more solutions
in the sigma model by allowing the integration constant in
eq.(\ref{arccos}) to be $(-K+i\zeta_0)$ instead of $(-K)$. For any real
$\zeta_0$ these configurations are real in Minkowski time. In analogy to
the L\"uscher-Schechter solutions, they possess a peculiar feature,
discovered in ref.\cite{Farhi} for $(3+1)$d Yang-Mills theory, that their
topological number, evaluated in Minkowski time, is non-integer in the
sense of ref.\cite{Farhi}. We will not discuss this property in the
present paper and concentrate on the case $\zeta_0=0$.

\section{The semiclassical process}

To interprete our solution in terms of a semiclassical process,
we have to evaluate the asymptotics of the fields $\Phi$ and $\bar{\Phi}$
which are
constructed according to eq.(\ref{FI}) from $w'$ and $\bar{w}'$ defined by
eq.(\ref{rot}). As discussed in Sect.2, their Fourier components at
$t\to+\infty$ are the amplitudes of the final coherent state, while their
anti-Feynman parts at $t=iT+t'$, $t'\to -\infty$ determine the amplitudes
of the initial coherent state. We consider the case $\psi_0\ll 1$ that
corresponds to the small number of initial particles.

\subsection{Final state}

We begin with the discussion of the final state. To the leading order in
$\psi_0$ (or in $w_0$, see eq.(\ref{w0})), the field at real time is
given by eq.(\ref{asymp}). At large $t$ it is a sum of waves moving
left and right along the light "cone". This field is weak, and to the
leading order in $w_0$
\[
\Phi=\sqrt{2}w'/g,~~~~~~~~~~~~~~
\bar{\Phi}=\sqrt{2}\bar{w'}/g
\]
 The Fourier components are easy to
calculate,
\[
	\alpha_k=\alpha^*_k=-\frac{\sqrt{8\pi}}{g}w_0\sqrt{k}\e^{-k},~~
	\beta_k=\beta^*_k=0,~~~~~\mbox{for}~k>0,
\]
\begin{equation}
  \alpha_k=\alpha^*_k=0,~~
	\beta_k=\beta^*_k=-\frac{\sqrt{8\pi}}{g}w_0\sqrt{-k}\e^{k},~~~~~
	\mbox{for}~k<0.
  \label{initial}
\end{equation}
We see that the final state contains particles moving right and
anti-particles moving left. The energy and total number of particles are

\begin{equation}
	E=\int_{-\infty}^{\infty}dk\mid k\mid(\alpha^*_k\alpha_k
	+\beta^*_k\beta_k)=\frac{4\pi}{g^2}w_0^2,
\label{energia}
\end{equation}
\begin{equation}
	n=\int_{-\infty}^{\infty}dk(\alpha^*_k\alpha_k
	+\beta^*_k\beta_k)=\frac{4\pi}{g^2}w_0^2.
\label{inumber}
\end{equation}
The typical momentum of a final particle is
\begin{equation}
       k_{final}\sim 1
\label{momentum}
\end{equation}
Of course, the latter  estimate has no direct physical significance, but
it will be useful for the comparison to the typical momentum of initial
particles.

\subsection{Initial state}

The calculation of the Fourier components of the fields $\Phi$ and
$\bar{\Phi}$ for the initial state is less straightforward.
At $w_0\ll 1$, the fields $w$
and $\bar{w}$ are large for  $\mbox{Im}~t$ close to $1$.
So, we really have to use the formalism developed in Sect.3.

Let us consider the regime $t=iT+t'$, $t'\to -\infty$ and concentrate on
the left part of the light cone where $x$ is close to $t'$. In this
region \[
  \zeta(x,t)=i\varphi=\frac{1}{2}\ln\frac{x_{-}-i}{x_{-}+i}
\]
where
\[
  x_{-}=x-t'-iT
\]
At $x_{-}$ far from the singularity, we may still use the approximation
(\ref{appr}) and obtain
\begin{equation}
  w=w_0\frac{x_-}{x_-+i}~,~~~\bar{w}=w_0\frac{x_-}{x_--i},
  \label{naive}
\end{equation}
 Notice that at $T\neq 0$ the fields $w$ and $\bar{w}$ are not
comlex-conjugate to
each other.

However, the approximation (\ref{appr}) is not valid when
$\zeta$ is close to the singularity $(K-iK')$
 which happens if $(1-T)$ is of order $w_0^2$ or  less and
$x_-$ near $-i$.
Let us evaluate $w$ and $\bar{w}$
for  these values of $\zeta$. Since $\zeta$ is close to $(K-iK')$, the most
suitable (and equivalent) form of the solution is as  follows,
\[
  w=\left(\frac{1+\sn(\zeta-K+iK',k_0)}{-1+\sn(\zeta-K+iK',k_0)}\right)^{1/2}
  \e^{i\varphi},
\]
where we made use of the transfomation formulae for elliptic functions (see
ref.\cite{GrRyzh}).
At small $\psi_0$,
 the leading-order approximation for the elliptic sine is $\sn~u=\tanh u$,
and we obtain
\begin{equation}
  w=\frac{-iw_0}{x_-+i}~,~~~\bar{w}=\frac{w_0}{2}~,
  \label{1st}
\end{equation}
This  is  consistent with eq.(\ref{naive}) when $x_-$ is close to $-i$, which
means that the "naive" formulae also work in the region near  the
singularity. However, to calculate $\Phi$ and $\bar{\Phi}$ we have to
find the
 derivatives of $w$ and $\bar{w}$
with respect to $x_-$. For $\partial_-w$ we can use either
eq.(\ref{naive}) or eq.(\ref{1st}), but
the leading-order expression obtained from eq.(\ref{1st}) for
$\partial_-\bar{w}$ vanishes, so we have to make use of
the second-order approximation\cite{GrRyzh},
\[
  \sn(u, k_0)=\left(1+\frac{k_1^2}{4}\right)\tanh\left[\left(1-\frac{k_1^2}{4}
  \right)u\right],
\]
from which we obtain for $x_-$ near $-i$,
\[
  \bar{w}=\frac{w_0}{2}\left(1+\frac{w_0^2}{2}\ln\frac{x_-+i}{-2i}
  +\frac{iw_0^4}{8}\frac{1}{x_-+i}+\frac{i}{2}(x_-+i)\right),
\]
and
\[
  \partial_-{\bar{w}}=\frac{iw_0}{4}\frac{(x_-+i-iw_0^2/2)^2}{(x_-+i)^2}.
\]
These expressions do not coincide with
eq.(\ref{naive}), so we have two approximate formulae for
$\partial_-\bar{w}$ in
two different regions, $x_-+i\sim w_0^2$ and
 $\mid x_-+i\mid\gg w_0^2$. One can
construct a function that approximates $\partial_-\bar{w}$ uniformly
in the two regions, namely,
\[
  \partial_-\bar{w}=-iw_0\frac{(x_-+i-iw_0^2/2)^2}{(x_--i)^2(x_-+i)^2}.
\]
This expression can now be used for calculating $\Phi$ and $\bar{\Phi}$.
After performing the rotation,
eq.(\ref{rot}),
we get
\[
  \partial_-\Phi=\frac{\sqrt{2}}{g}\frac{\partial_-w'}{1+\bar{w}'w'}=
  \frac{\sqrt{2}}{g}
  \frac{1+w_0\bar{w}}{1+w_0w}\frac{\partial_-w}{1+\bar{w}w}=
\]
\[
  \frac{i\sqrt{2}}{g}w_0\frac{1}{(x_-+i-iw_0^2/2)(x_-+i-iw_0^2)},
\]
\[
  \partial_-\bar{\Phi}=\frac{\sqrt{2}}{g}
  \frac{\partial_-\bar{w}'}{1+\bar{w}'w'}=
\]
\[
  \frac{(x_-+i-iw_0^2/2)(x_-+i-iw_0^2)}{(x_--i)^2(x_-+i)^2},
\]

The calculation of the Fourier components is now straightforward. We find
for the anti-Feynman parts
\[
	\alpha_k=0,
\]
\begin{equation}
	\beta_k=\frac{1}{g}\sqrt{\frac{\pi}{8}}w_0^5
	\frac{1}{\sqrt{k}}\left(k-\frac{3}{w_0^2}\right)\e^{-\epsilon k},
\label{anti}
\end{equation}
and for the Feynman parts
\[
	\bar{\alpha}_k=-\frac{\sqrt{8\pi}}{g}w_0\sqrt{k}
	\e^{-(2-\epsilon)k},
\]
\begin{equation}
	\bar{\beta}_k=\frac{1}{g}\sqrt{32\pi}\frac{1}{w_0}\frac{1}{\sqrt{k}}
  \left(\e^{-(w_0^2/2-\epsilon)k}-\e^{-(w_0^2-\epsilon)k}\right)
  \label{final}
\end{equation}
where
\[
\epsilon=1-T\ll 1
\]
 is yet an arbitrary parameter in the interval $(w_0^6/32,
w_0^2/2)$. Eqs.(\ref{anti}),(\ref{final}) are valid for the right-moving waves
($k>0$), the expressions for the left-moving waves  ($k<0$) are completely
analogous.

  As pointed out in Sect.2, the contour in the complex time plane, i.e,
the value of the parameter $\epsilon$, should
be chosen in such a way that the initial coherent state has the same
average energy as the final one. In our case this means
\begin{equation}
	\int_{-\infty}^{\infty}dk\mid k\mid(\alpha^*_k\alpha_k
	+\beta^*_k\beta_k)=\frac{4\pi}{g^2}w_0^2,
\label{cond'n}
\end{equation}
Recall that $\alpha^*_k$ and $\beta^*_k$ are complex congugate to
$\alpha_k$ and $\beta_k$, respectively, and are {\it not} equal to the
Feynman parts of the field.
Eq.(\ref{cond'n}) gives
\[
 \epsilon=w_0^{8/3}/4
\]
Eq.(\ref{anti}) then implies that the typical momentum of an incoming
particle is
\[
   k_{initial}\sim w_0^{-8/3}
\]
which is much larger than the typical momentum of an outgoing particle,
eq.(\ref{momentum}).
  The number of particles in
the initial state is much smaller than the number of final particles,
\begin{equation}
  n=\int_{-\infty}^{\infty} dk(\alpha^*_k\alpha_k+\beta^*_k\beta_k)=
  \frac{\pi}{g^2}w_0^{14/3}.
\label{fnumber}
\end{equation}
Eq.(\ref{fnumber}) shows that the process with small number of incoming
particles indeed corresponds to the limit $w_0\to0$. In this limit the
number of outgoing particles is also small, see eq.(\ref{inumber}), so that
\begin{equation}
  n_{final}=\frac{1}{g^2}(g^2 n_{initial})^{3/7}
\label{estimate}
\end{equation}
Notice that the number of final particles is non-analytic in
 $\nu=g^2 n_{initial}$.

Finally, we note that our solution does not correspond to any "microcanonical
process" in the sense of Sect.2. Indeed, the ratio
$\bar{\beta}_k/\beta^*_k$ is
not a positive definite function of momentum, which would be necessary for
the interpretation of the solution in terms of the "microcanonical"
probability.

\subsection{Transition probability}

The probability of the process described by our solution is calculated
according to eq.(\ref{prob}).
We have seen in Sect.4 that the imaginary part of the action for our
configuration is exactly equal to the instanton action.
The parameter $T$ at small $w_0$ is
\[
T=1+O(w_0^{8/3})
\]
 and eqs.(\ref{energia}),(\ref{anti}),(\ref{final}) and (\ref{fnumber})
give for various terms in the exponent
\[
  2ET=\frac{8\pi w_0^2}{g^2}
\]
\[
  \int dk(\bar{\alpha}_k \alpha_k+\bar{\beta_k}\beta_k)_{initial}
  = (1-3\ln 2)\frac{4\pi w_0^2}{g^2}
\]
and
\[
  -\int dk(\alpha^*_k \alpha_k+\beta^*_k\beta_k)_{initial} = O(w_0^{14/3})
\]
So, we have for the transition probability
\[
\sigma=\exp\left(-\frac{8\pi}{g^2}+\frac{12\pi(1-\ln2)}{g^2}w_0^2
          +O(w_0^{14/3})\right)
\]
Equivalently, the transition probability may be expressed through the
number of the incoming particles,
\begin{equation}
  \sigma=\exp\left( -\frac{8\pi}{g^2}+
    12(1-\ln2)\left(\frac{\pi}{g^2}\right)^{4/7}(n_{initial})^{3/7}\right)
\label{est}
\end{equation}
Clearly, at small $n_{initial}$, the transition probability is suppressed
by the instanton exponent.

 \section{Conclusions}

In this paper we have found an analytic solution to the classical field
equations in the massless $O(3)$ sigma model in $(1+1)$ dimensions. This
solution describes a multi-particle instanton-like transition at high
energy. Most properties of this solution agree with the earlier
expectations\cite{RST} concerning the structure of its singularities in
the complex time plane and the behavior in the euclidean domain in the
limit of small number of incoming particles. We have found, however, that
this solution does {\it not} correspond to the maximum transition
probability among  all initial states with given number of paricles and
given energy (or any "microcanonical" probability in general).

Unless the limit $\nu=g^2 n_{initial} \to 0$ is very sensitive to the
details of the structure of the initial state (which seems rather
unlikely), our results indicate that well above the sphaleron, the
instanton-induced cross section again becomes suppressed by the instanton
exponent, and the number of outgoing particles is again
much smaller than $\mbox{const}/g^2$. However, on
the basis of these results one may anticipate, for $2\to\mbox{any}$
process, the non-analiticity in $g^2$ of both the number of outgoing
particles and the exponent for the total cross section. Indeed, if we
naively set $n_{initial}\sim 1$ in eqs.(\ref{estimate}) and (\ref{est}),
we would get
\[
       n_{final}\sim \frac{1}{g^{8/7}}
\]
 and
\[
\sigma=\left(-\frac{8\pi}{g^2}+\frac{\mbox{const}}{g^{8/7}}\right)
\]

To obtain the solution analytically, we heavily made use of specific
properties of the massless theory, in particular, its conformal
invariance. Therefore, we were unable to make contact to the perturbative
calculations about the instanton that are relevant at low energies.
Hopefully, the complete solution of this problem may be possible by
numerical methods.

The authors are indebted to E.Farhi, V.Khoze, L.McLerran, A.Mueller,
A.Ring\-wald, M.Shaposhnikov, A.Vainshtein and M.Voloshin for discussions.
 P.T. thanks CERN for hospitality and the Eloisatron project for partial
support, V.R. thanks TPI, University of Minnesota, for hospitality.

\newpage

\begin{picture}(300,200) \put
(50,100){\vector(1,0){200}} \put (150,40){\vector(0,1){130}} \thicklines \put
(70,140){\line(1,0){80}} \put (150,140){\line(0,-1){40}} \put
(150,100){\line(1,0){80}} \put (70,145){$A$} \put (135,145){$B$} \put
(185,105){$C$} \put (157,135){$T$} \put (155,165){Im~$t$} \put
(240,87){Re~$t$} \put (140,0){Fig.1} \end{picture}

\newpage

 \begin{picture}(300,200) \put (0,100){\vector(1,0){250}} \put
(125,0){\vector(0,1){150}} \bezier {200}(0,50)(25,150)(50,50) \bezier
{200}(50,50)(75,-50)(100,50) \bezier {200}(100,50)(125,150)(150,50) \bezier
{200}(150,50)(175,-50)(200,50) \bezier {200}(200,50)(225,150)(250,50) \put
(143,75){\line(1,0){64}} \put (143,75){\line(0,1){25}} \put
(143,100){\circle*{1}} \put (143,105){$\psi_0$} \put
(130,145){$U(\psi)=-(1-\cos\psi)$} \put (245,105){$\psi$} \put
(120,-25){Fig.2} \end{picture}

\newpage

\begin{picture}(400,400) \put
(200,50){\vector(1,0){200}} \put (200,50){\vector(0,1){250}} \put
(200,50){\vector(-4,-1){200}}
\put (0,200){\line(4,1){400}} \put (0,300){\line(4,-1){400}}
\thicklines
\put (0,280){\line(4,-1){160}} \put (0,180){\line(4,1){160}}
\bezier {200}(160,240)(200,230)(160,220)
\put (0,250){\line(4,-1){160}} \put (0,150){\line(4,1){160}}
\bezier {200}(160,210)(200,200)(160,190)
\thinlines
\put (100,130){\vector(0,1){40}} \put (70,115){poles}
\put (350,160){\vector(0,1){45}} \put (250,145){essential singularities}
\put (205,295){Im~$t$} \put (390,37){Re~$t$} \put (0,10){$x$} \put
(205,255){1} \put (190,-25){Fig.3} \end{picture}


\begin{thebibliography}{99}

\bibitem{Ringwald}
        A.Ringwald, {\em Nucl.Phys.} {\bf B330} (1990) 1.
\bibitem{Espinosa}
        O.Espinosa, {\em Nucl.Phys.} {\bf B334} (1990) 310.
\bibitem{MVV}
        L.McLerran, A.Vainshtein and M.B.Voloshin, {\em Phys. Rev.}
        {\bf D42} (1990) 171.
\bibitem{Zakharov}
    V.I.Zakharov,  {\em Classical Corrections to Instanton-Induced
Amplitudes}, preprint TPI-MINN-90/7-T, 1990;\\
{\em Nucl.Phys.}, {\bf B371}
(1992) 637.
 \bibitem{KRT}
       S.Yu.Khlebnikov, V.A.Rubakov and P.G.Tinyakov, {\em
     Mod.Phys.Lett.} {\bf A5} (1990) 1983;\\ {\em Nucl.Phys.}
     {\bf B350} (1991) 441.
 \bibitem{Porrati}
       M.Porrati, {\em Nucl.Phys.} {\bf B347} (1991) 371.
\bibitem{ArnoldMattis}
       P.B.Arnold and M.P.Mattis, {\em Phys.Rev.} {\bf D42} (1990) 1738.
\bibitem{Yaffe}
       L.Yaffe, {\em Scattering Amplitudes in Instanton Background}. In:
M.Mattis and E.Mottola, eds., {\em Proc. of the Santa Fe
Workshop}, World Scientific, Singapore, 1990.
\bibitem{Mueller1}
       A.Mueller, {\em Nucl.Phys.} {\bf B348} (1991) 310.
\bibitem{Mueller2}
       A.Mueller, {\em Nucl.Phys.} {\bf B353} (1991) 44.
\bibitem{KhlebnikovTinyakov}
       S.Yu.Khlebnikov and P.G.Tinyakov, {\em Phys.Lett} {\bf 269B}
       (1991) 149;\\
       P.B.Arnold and M.P.Mattis, {\em Phys.Rev.} {\bf D44} (1991) 3650;\\
       M.P.Mattis, L.McLerran and L.G.Yaffe, {\em Phys.Rev.} {\bf D45}
       (1992) 4294.
\bibitem{RubakovTinyakov}
  V.A.Rubakov and P.G.Tinyakov, {\em Phys. Lett.} {\bf B279} (1992) 165.
\bibitem{Tinyakov}
  P.G.Tinyakov, {\em Phys. Lett.} {\bf B284} (1992) 410.
\bibitem{RST}
  V.A.Rubakov, D.T.Son and P.G.Tinyakov,
  {\em Phys.Lett} {\bf 287B} (1992) 342.
\bibitem{Mueller3}
    A.Mueller, {\em Comparing Two Particle and Multi-Particle Initiated
Processes in the One Instanton Sector}, preprint CU-TP-572, 1992.
\bibitem{Mottola}
      E.Mottola and A.Wipf, {\em Phys.Rev.} {\bf D39} (1989) 588.
\bibitem{KRTold}
      S.Yu.Khlebnikov, V.A.Rubakov and P.G.Tinyakov, {\em Nucl.Phys.}
      {\bf B347} (1990) 783.
\bibitem{Dorey}
       N.Dorey, {\em Multi-Instanton Valleys in the $O(3)$ Sigma-Model},
         preprint LA-UR-92-1380, 1992.
\bibitem{KRTperiodic}
     S.Yu.Khlebnikov, V.A.Rubakov and P.G.Tinyakov, {\em Nucl.Phys.}
      {\bf B367} (1991) 334.
\bibitem{British}
     J.Bossart and C.Weisendanger, {\em Phys.Rev.} {\bf D46} (1992) 1820.
\bibitem{Luscher}
     M.L\"uscher, {\em Phys.Lett} {\bf 70B} (1977) 321.
\bibitem{Schechter}
     B.Schechter, {\em Phys.Rev.} {\bf D16} (1977) 3015.
\bibitem{Farhi}
     E.Farhi, V.V.Khoze and R.Singleton,Jr., {\em Minkowski Space
Non-Abelian Classical Solutions with Non-Integer Topological
Number}, MIT preprint, 1992.
\bibitem{GrRyzh}
{\em Handbook of Mathematical Functions},  M.Abramowitz and
I.Stegun, eds., Dover, N.-Y.,1964.

I.S.Gradstein and I.M.Ryzhik, {\em Table of Integrals, Series and Products},
               Academic Press, N.-Y., 1980.
\end{thebibliography}
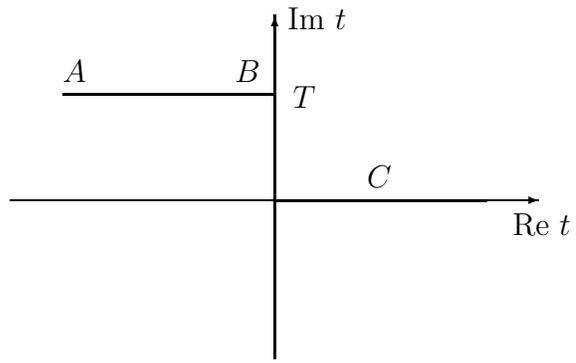
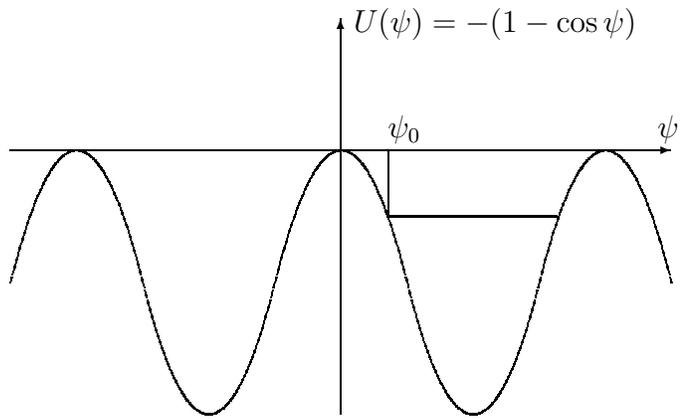
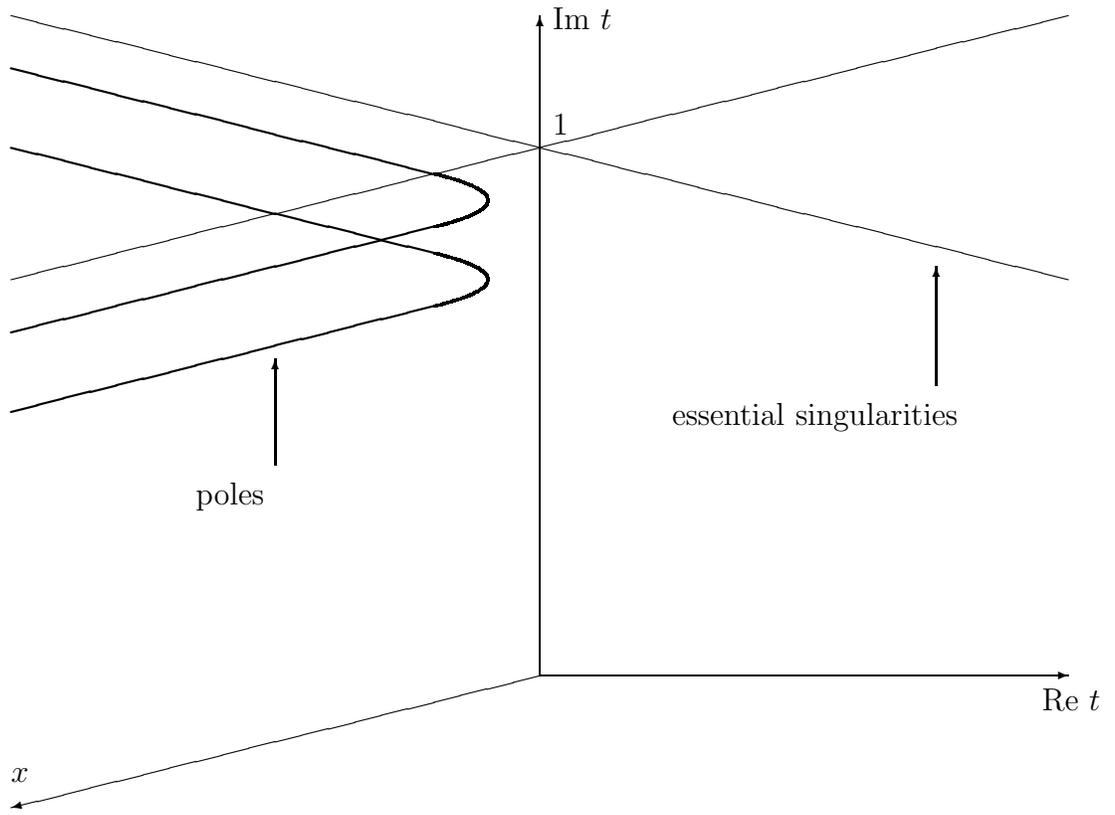
\end{document}